\documentclass[
   ,final            
  ]
  {aipproc}
\usepackage{graphics}
\usepackage{subfigure}
\layoutstyle{8x11single}

\begin{document}

\title{Gas dynamics in strong centrifugal fields}


\author{S.V. Bogovalov, V.A. Kislov and I.V. Tronin}{
  address={National research nuclear university ``MEPhI'', Kashirskoje shosse, 31,115409, Moscow, Russia}
}



\begin{abstract}
Dynamics of waves generated by scopes in gas centrifuges (GC) for isotope separation is considered. The centrifugal acceleration in the GC
reaches values of the order of $10^6$g. The centrifugal and Coriolis forces modify essentially  the conventional sound waves. Three families of the waves with different polarisation and dispersion exist in these conditions. Dynamics of the flow in the model GC Iguasu is investigated numerically. Comparison of the results of the numerical modelling of the wave dynamics with the analytical predictions is performed. New phenomena of the resonances in the GC is found. The resonances occur for the waves polarized along the rotational axis having the smallest dumping due to the viscosity. 
\end{abstract} 

\maketitle


\section{Introduction}
Strong centrifugal fields of the order of $10^6$g are used in GC for the isotope separation. Gas in GC does not trivially corotates with the rotor of the GC. Secondary axial flows are excited  in the GC intentionally to amplify the efficiency of the isotope separation. One of the method of  the creation of the secondary flows is the local mechanical brake of the gas  by scopes for the leakage of the gas from the GC.  The working gas in the GC rotates with the velocity exceeding the sound velocity on the factor 5-6. Therefore, interaction of the gas with the scopes is accompanied by the formation of the shock waves propagating in the axial direction.  The impact of these waves on the separative efficiency of the gas centrifuges is not investigated. Moreover, even the characteristics of the linear waves in the condition of the strong centrifugal fields are still  unknown \cite{rotation}. 
In this work we discuss the basic characteristics of the linear waves in the rotating gas  in the axisymmetric approximation. The problem of an unsteady flow in the axisymmetrically rotating gas is solved analytically. The same problem is solved numerically for the simplest model GC consisting of two cameras: upper camera where a periodically varying force affects the gas and working camera. 

The problem of numerical modelling of the gas dynamics in strong centrifugal fields reaching $10^6$ g is of special interest. Conventional CFD codes give unphysical results in these conditions on two basic reasons. Firstly, the strong centrifugal field produces unphysical mass flows at the interfaces between the control volumes. Secondly, the velocity of the secondary axial flows in gas centrifuges are of the order of centimetres per seconds, in some cases even millimetres per seconds. To resolve accurately such a velocity on the background of rotational velocity of the order of 600 meters per second is not trivial problem at the moderate mesh resolution. The group from  National research nuclear university (MEPHI) has developed a numerical code based on the modified Rhie-Chow interpolation scheme \cite{rhie_chow,bogtr} to investigate gas dynamics in these conditions focusing on the propagation of the waves in GC.

\section{Characteristics of the linear waves}

Analytical analysis of the characteristics of the waves in strong centrifugal fields is performed in dissipationless approximation. The viscosity and thermal conductivity of the gas have been neglected.  Linearisation of the full system of equations defining dynamics of the gas in the rotating cylindrical frame system gives 
\begin{equation}
 \frac{\partial \bar{\rho}}{\partial t}+\frac{\partial \left(r \rho_0 \bar{v_r}\right)}{r \partial r}+\frac{\partial \rho_0 \bar{v_z}}{\partial z}=0,
\end{equation}
\begin{equation}
 \rho_0 \frac{\partial \bar{v_r}}{\partial t}-2 \rho_0 \omega \bar{v_\phi}-\bar{\rho}\omega^2 r=-\frac{\partial \bar{p}}{\partial r},
\end{equation}
\begin{equation}
 \rho_0 \frac{\partial \bar{v_\phi}}{\partial t}+2\rho_0 \omega \bar{v_r}=0,
\end{equation}
\begin{equation}
 \rho_0\frac{\partial \bar{v_z}}{\partial t}=-\frac{\partial \bar{p}}{\partial z},
\end{equation}
\begin{equation}
 \rho_0 c_p \frac{\partial \bar{T}}{\partial t}=\frac{\partial \bar{p}}{\partial t}+\rho_0 \omega^2 r \bar{v_r},
\end{equation}
where all the terms with upper bar are deviations from the equilibrium state corresponding to the rigid body rotation of the gas. $\rho_0$  is the equilibrium density
 \begin{equation}
 \rho_0=\rho_w exp(\frac{M \omega^2}{2RT_0}\left( r^2-a^2\right)),
\end{equation}
$\omega$ - is the angular velocity of the rotor of the GC, $M$ - molar mass of the gas, $\rho_w$ - density at the wall of the rotor, $a$- radius of the rotor, $T_0$ - temperature of the gas.  
Let us assume that periodical along axial direction $z$ wave has the wave vector $k$ and frequency  $\Omega$.  All the perturbations take a form
\begin{eqnarray}
 \bar{v_r}=V_r e^{-i \Omega t+i k z}, \bar{v_\phi}& = &V_\phi e^{-i \Omega t+i k z}, \bar{v_z}=V_z e^{-i \Omega t+i k z},\nonumber\\ 
\bar{p}=p e^{-i \Omega t+i k z}, \bar{\rho}& = &\rho e^{-i \Omega t+i k z},\bar{T}=T e^{-i \Omega t+i k z}.\nonumber
\end{eqnarray}
Here $\Omega$ - frequency of the wave, $k$ - wave vector along axial direction. 
\begin{figure}
\centering
\includegraphics[width=0.7\linewidth]{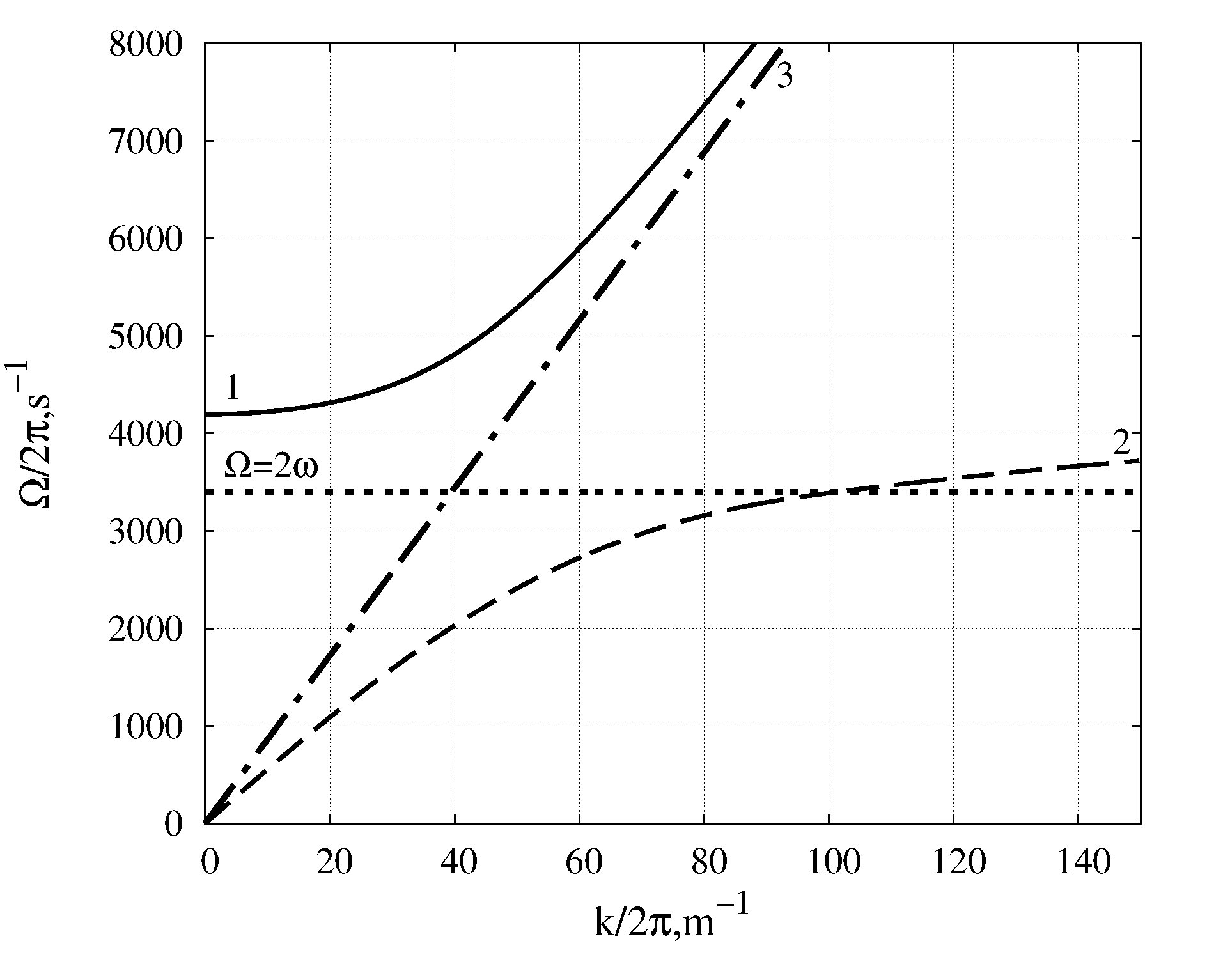}
\caption{Law of dispersion for three families of the waves in the GC. Only curves corresponding to the radial ground states are shown for the families 1 and 2. Curve 3 corresponds to the waves polarized along the rotational axis. They have the law of dispersion $\Omega = kc$}
\label{fig:dispersion}
\end{figure}
Solution of the system of equations gives three families of waves. Dispersion curves for the waves are shown in fig. 1. They were calculated for the parameters of GC 
given in table. 1. Dispersion curve of the first family of the waves is located above the frequency of the conventional inertial waves \cite{rotation} and above the dispersion curve of the conventional sound waves $\omega=ck$. All the perturbations of the velocities in these waves are concentrated near the axis of rotation. 

Dispersion curves of the 
second family are located below the line  $\omega=c k$.  Perturbations of the flow from these waves are also concentrated close to the rotational waves. 

The waves of the first and second family are polarised in all $r$, azimuthal and $z$ directions.  

Dispersion curve of third family coincides with the line  $\omega=ck$. These waves are polarized exactly along the rotational axis. The azimuthal velocity $V_z$ is connected with the perturbation of the pressure as follows 
\begin{equation}
 V_z=\frac{k p}{\Omega \rho_0},
\label{eqvz}
\end{equation}
while the perturbation of pressure has rather trivial form
\begin{equation}
 p=\bar p_w \exp{\frac{\omega^2(r^2-a^2)}{2c^2}},
\end{equation}
where $\bar p_w$ is the perturbation of the pressure at the wall of the rotor. Detailed discussion of the characteristics of these waves is presented in \cite{p1}.

The last family of the waves is the most interesting for the physics of GC.  The deviations of the velocity, density and temperature for the waves of the first and second families are located close to the axis of rotation. They have very short propagation length due to viscose dumping. The waves  of the third family have almost uniform distribution of the deviations on the radius. They will have a smallest dumping and are able to propagate at larger distance. It is possible to predict for these waves a phenomena   of resonances. Because of large transport length these waves can reflect from the end caps of GC many times. The GC can operate like a resonator. The intensity of the waves can increase essentially when the doubled length of the GC equals to an integer number of wavelengths. To check this assumption  numerical simulation has been performed.

\begin{table}
\caption{\label{tab:1}Basic parameters of GC Iguasu}
\begin{tabular}{|c|c|}
\hline
Parameter & Value  \\
\hline
$M$ & 352 g/mol\\
\hline
$a$ & 0.065 m \\
\hline
$\omega$ & $2\pi\times 1700 s^{-1}$ \\
\hline
$T_0$ & 300 K\\
\hline
$p_w$ & 80 mm Hg\\
\hline
$c_p$& 385 $J\cdot K/kg$ \\
\hline
$c$& 86 m/s\\
\hline
\end{tabular}
\end{table}
\subsection{Numerical results}

For investigation of dynamics of the waves a software for modelling of the unsteady flows in the strong centrifugal fields has been developed by the group in MEPHI. 
The results of the verification of this software have been presented in \cite{jvm,cf}.  

Geometry of the computational domain is shown in fig. 2 (left panel). The problem is solved in axisymmetric approximation. The GC consists of two chambers. Upper chamber with the height  2 cm contains a small region where  periodical in time braking force  takes place. Strong heating of the gas results into shock waves propagating through the inner hole of a baffle. The waves propagate along the rotational axis in the working camera with length 20 cm, reflect from the lower end cap and return back. Distribution of the pressure in one of the time moment is shown in fig. 2 (left panel). An analysis shows that the waves propagating in the working chamber are the waves of third family. All other waves are dumped at smaller distances. 

\begin{figure}
\centering
\begin{minipage}[h]{0.49\linewidth}
\centering
\includegraphics[width=0.5\linewidth]{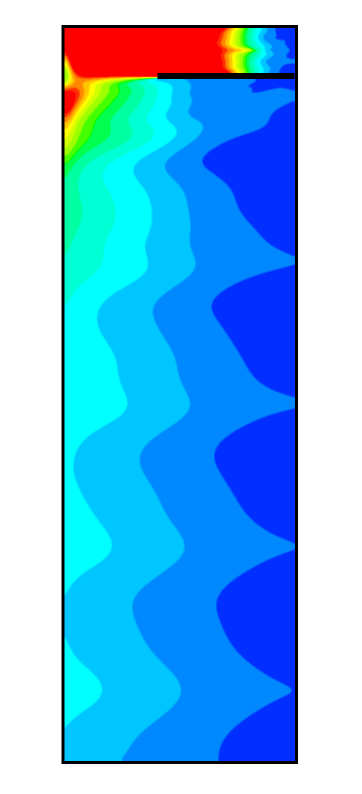}
\caption{first}
\end{minipage}
\hfill
\begin{minipage}[h]{0.55\linewidth}
\centering
\includegraphics[width=1.\linewidth]{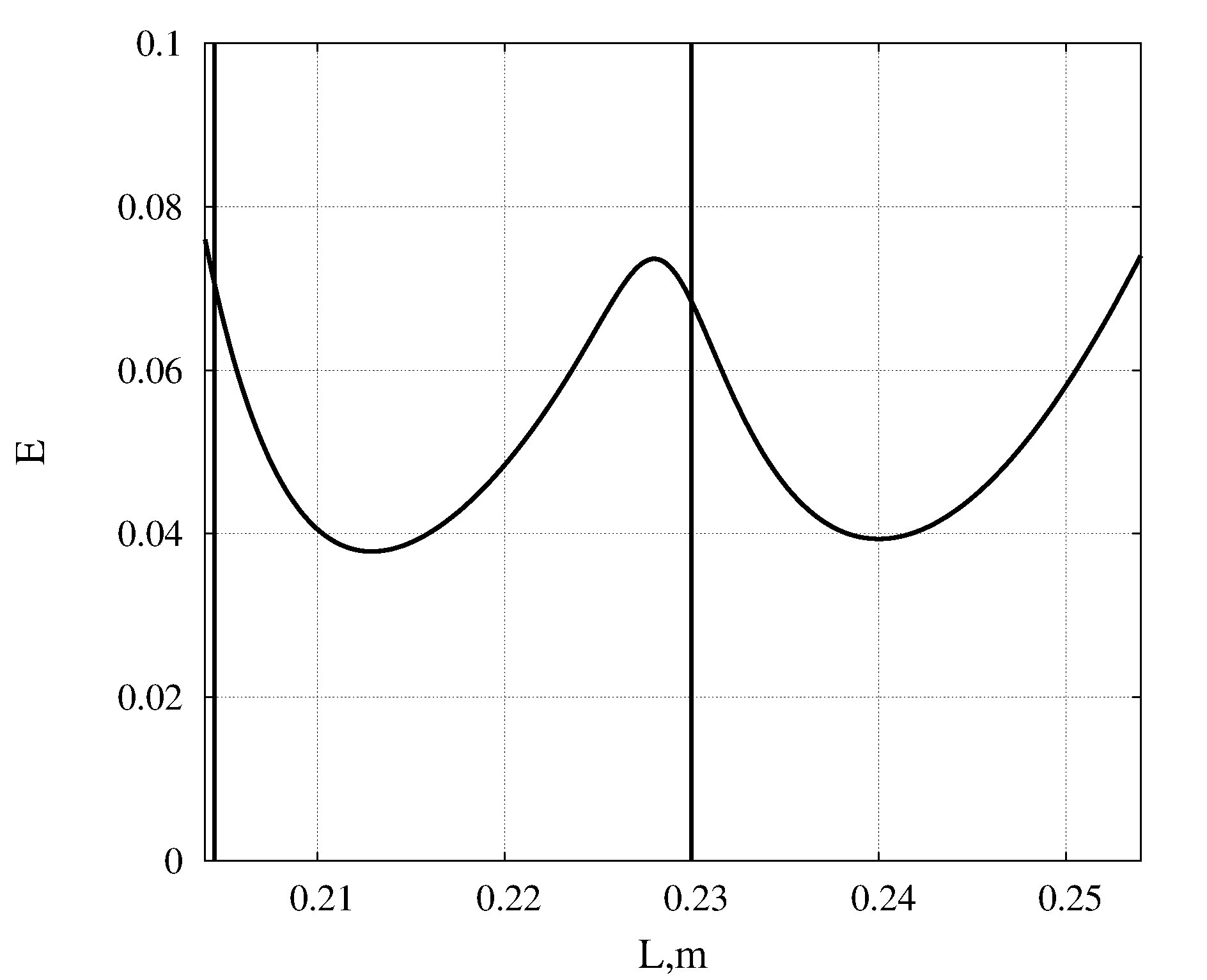}\\
\caption{Wave generation in the model centrifuge. Snapshot of the pressure in the computational domain is shown in the left panel. Dependence  of the averaged over the computational domain and time intensity of the waves on the length of the GC is shown in the right panel.}
\end{minipage}
\end{figure}

The resonance occurs at the length of the working chamber $L$ satisfying to the condition $L={\lambda\over 2}n$, where $\lambda$ is the wave length and $n$ is an integer number. Fig. 2(right panel) shows the averaged amplitude of the velocity of plasma over the working chamber in dependence on the length $l$. The resonances are  well  visible. The figure shows as well that resonances repeat at the GC length variation equal to the half of wavelength.  The impact of the resonances on the separative efficiency of GC would be the most interesting  and important  next step in investigation of the gas dynamics in strong centrifugal fields.

\section{Conclusion}

The analysis of the problem of the wave propagation in the gas placed in strong centrifugal field shows that three type of waves with different polarization are possible in GC. Two types should have strong dumping due to the molecular viscosity. The waves polarized along the rotational axis has the smallest dumping. Numerical solution of the non stationary problem of the wave dynamics in the GC shows that indeed these waves are produced at the periodical braking of the rotating gas. It is interesting that the slow dumping of these waves results into the phenomena of resonance when natural number of the half waves equals to the length of the working camera. In this case the averaged energy of the waves increases a few times. It is reasonable to guess that such an amplification of the amplitudes of the waves will affect the separation of the isotopes.       

\begin{theacknowledgments}
 The work has been performed under support of state order of Ministry of education and science of Russia, grant no. 3.726.2014/K
\end{theacknowledgments}
\bibliographystyle{aipproc} 

\begin{thebibliography}{9}
\bibitem{rotation} H.P. Greenspan, \emph{The Theory of Rotating Fluids}, Cambridge Univ. Press., 1968
\bibitem{rhie_chow} C.M. Rhie, W.L. Chow, \emph{AIAA Journal}, 1983, Vol. 21, pp. 1525-1532
\bibitem{bogtr} S. V. Bogovalov, I.V. Tronin, \emph{Computational Mathematics and Mathematical Physics}, 2014, in preparation
\bibitem{p1} S.V.Bogovalov, V.A.Kislov, I.V.Tronin. \emph{J.Fluid Mech.}, 2014, submitted
\bibitem{jvm} V. I. Abramov, S. V. Bogovalov, V. D. Borisevich,  et al. \emph{Computational Mathematics and Mathematical Physics}, 2013, Vol. 53, No. 6, pp. 789-797.
\bibitem{cf}  S. V. Bogovalov, V. D. Borisevich, V.D., Borman  et al. \emph{Computers \& Fluids}, 2013, 86, pp. 177-184.




\end{thebibliography}



\end{document}